\documentclass{iaus}
\usepackage{graphicx}

\newcommand{\ie}{i.e., }

\newcommand{\Msun}{M_{\odot}}

\newcommand{\ergs}{ergs~s$^{-1}$}

\newcommand{\Nifs}{$^{56}$Ni}

\newcommand{\Ed}{\dot{E}_{\rm dep}}
\newcommand{\Edep}{\dot{E}_{\rm dep,51}}

\newcommand{\Mms}{M_{\rm ms}}
\def\gsim{\mathrel{\rlap{\lower 4pt \hbox{\hskip 1pt $\sim$}}\raise 1pt
\hbox {$>$}}}
\def\lsim{\mathrel{\rlap{\lower 4pt \hbox{\hskip 1pt $\sim$}}\raise 1pt
\hbox {$<$}}}

\newcommand{\Mbh}{M_{\rm rem}}
\newcommand{\Et}{{E}_{\rm dep}}

\newcommand{\thj}{\theta_{\rm jet}}
\newcommand{\fth}{f_{\rm th}}
\newcommand{\Gj}{\Gamma_{\rm jet}}

\newcommand{\Min}{M_0}
\newcommand{\Rin}{R_0}
\newcommand{\Mfe}{M{\rm (Fe)}}

\title[SN Nucleosynthesis in the Early Universe]
{Supernova Nucleosynthesis in the Early Universe}

\author[N. Tominaga et al.]
{Nozomu Tominaga$^1$, Hideyuki Umeda$^2$, Keiichi Maeda$^3$, \\
Ken'ichi Nomoto$^{3,2}$, \and Nobuyuki Iwamoto$^4$
}

\affiliation{$^1$Optical and Infrared Astronomy Division, National
Astronomical Observatory, Mitaka, Tokyo, Japan
\\ email: {\tt nozomu.tominaga@nao.ac.jp} \\[\affilskip]
$^2$Department of Astronomy, School of Science,
University of Tokyo, Bunkyo, Tokyo, Japan \\[\affilskip]
$^3$Institute for the Physics and Mathematics of the Universe, 
University of Tokyo, Kashiwa, Chiba, Japan \\[\affilskip]
$^4$Nuclear Data Center, Nuclear Science and Engineering
Directorate, Japan Atomic Energy Agency, Tokai, Ibaraki, Japan}

\pubyear{2008}
\volume{255} 
\pagerange{1--5}
\jname{Low-Metallicity Star Formation: From the First Stars to Dwarf Galaxies}
\editors{L.K. Hunt, S. Madden \& R. Schneider, eds.}
\begin{document}

\maketitle

\begin{abstract}
 The first metal enrichment in the universe was made by supernova (SN) 
 explosions of population (Pop) III stars. The trace remains in
 abundance patterns of extremely metal-poor (EMP) stars. We 
 investigate the properties of nucleosynthesis in Pop III SNe by means
 of comparing 
 their yields with the abundance patterns of the EMP stars. We focus on 
 (1) jet-induced SNe with various 
 energy deposition rates [$\Ed=(0.3-1500)\times10^{51}$\ergs], and (2) 
 SNe of stars with various main-sequence
 masses ($\Mms=13-50\Msun$) and explosion energies
 [$E=(1-40)\times10^{51}$ergs]. The 
 varieties of Pop III SNe can explain varieties of the EMP stars: 
 (1) higher [C/Fe] for lower [Fe/H] and (2) trends of
 abundance ratios [X/Fe] against [Fe/H]. 
\end{abstract}

\firstsection
\section{Introduction}
\label{sec:intro}

Long-duration $\gamma$-ray bursts (GRBs) have been found to be
accompanied by luminous and energetic Type Ic supernovae [SNe Ic, called
hypernovae (HNe)] (\eg \cite[Galama \etal\ 1998]{gal98}). 
Although the explosion mechanism is still
under debate, photometric observations (a ``jet break'', \eg
\cite[Frail \etal\ 2001]{fra01}) and spectroscopic observations (a
nebular spectrum, \eg \cite[Maeda \etal\ 2002]{mae02})
indicate that they are aspherical explosions with jet(s).

The aspherical explosions are also indirectly suggested from the abundance patterns
of extremely metal-poor (EMP) stars with [Fe/H] $<-3$.\footnote{Here [A/B] 
$\equiv\log_{10}(N_{\rm A}/N_{\rm B})-\log_{10}(N_{\rm A}/N_{\rm B})_\odot$,
where the subscript $\odot$ refers to the solar value and $N_{\rm A}$
and $N_{\rm B}$ are the abundances of elements A and B, respectively.}
The EMP stars are suggested to show nucleosynthesis yields
of a single core-collapse SN (\eg
\cite[Beers \& Christlieb 2005]{bee05}). Particularly, C-enhanced EMP (CEMP) stars have
been well explained by the faint SNe with large fallback (\cite[Umeda \&
Nomoto 2005; Iwamoto \etal\ 2005; Nomoto \etal\ 2006; Tominaga \etal\
2007b]{ume05,iwa05,nom06,tom07b}). On the other hand, some CEMP stars
show enhancement of Co and Zn (\eg \cite[Depagne \etal\ 2002]{dep02})
that requires explosive nucleosynthesis under high
entropy. In a {\sl spherical} model, however, a high entropy explosion
is equivalent to a
high energy explosion that inevitably synthesizes a large amount of
\Nifs, \ie leads a bright SN (\eg \cite[Woosley \& Weaver
1995]{woo95}). This incompatibility will be solved if a faint
SN is associated with a narrow jet within which a high entropy region
is confined (\cite[Umeda \& Nomoto 2005]{ume05}). 

\section{Models}
\label{sec:model}

We investigate the jet-induced explosions (\eg \cite[Maeda \& Nomoto
2003; Nagataki \etal\ 2006]{mae03,nag06}) of
$40\Msun$ Pop III stars (\cite[Umeda \& Nomoto 2005; Tominaga \etal\
2007b]{ume05,tom07b}) using a two-dimensional
special relativistic Eulerian hydrodynamic code (\cite[Tominaga 2009]{tom07c}). 

\begin{figure}
\begin{center}
 \includegraphics[width=1.7in]{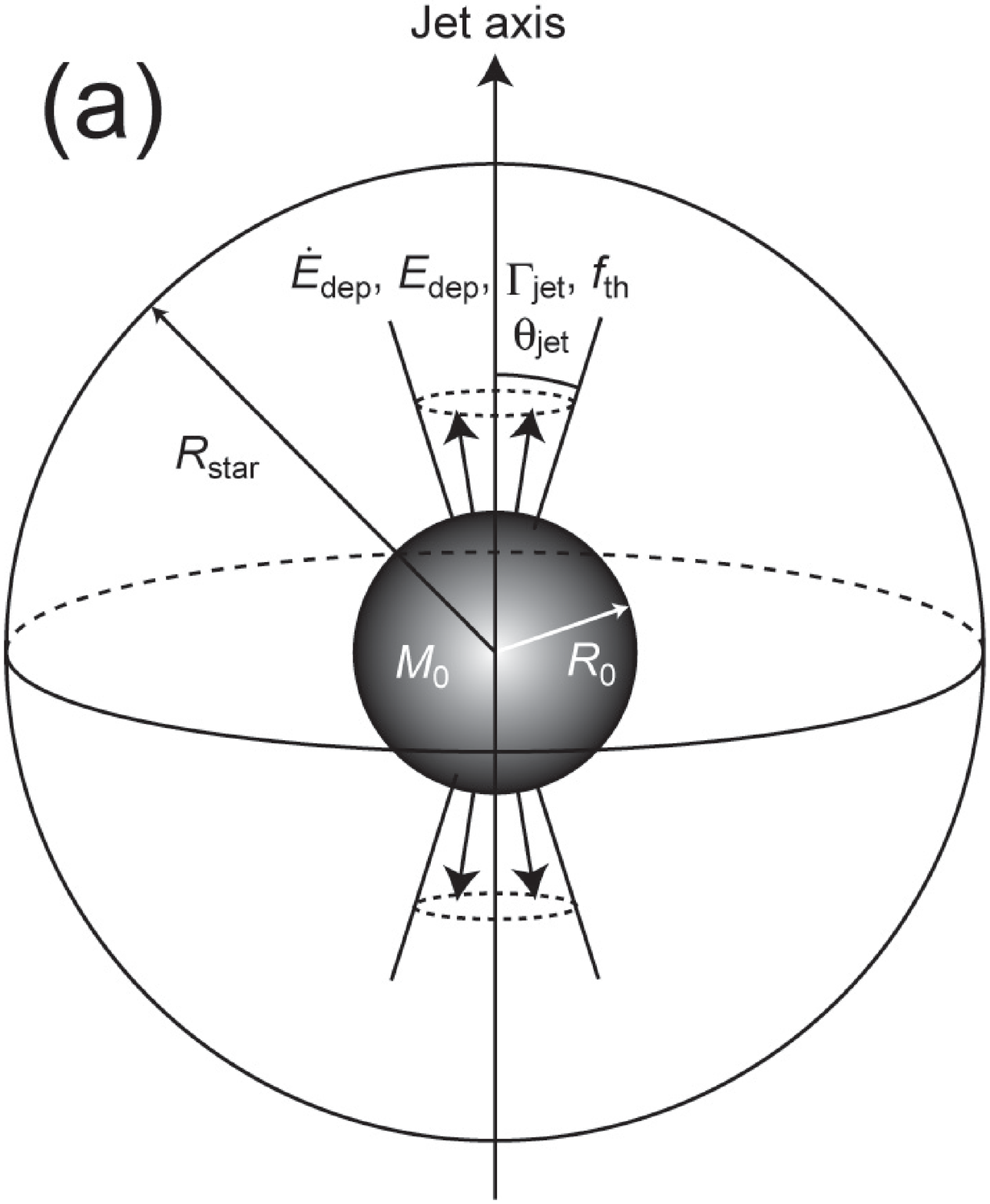} 
 \includegraphics[width=1.7in]{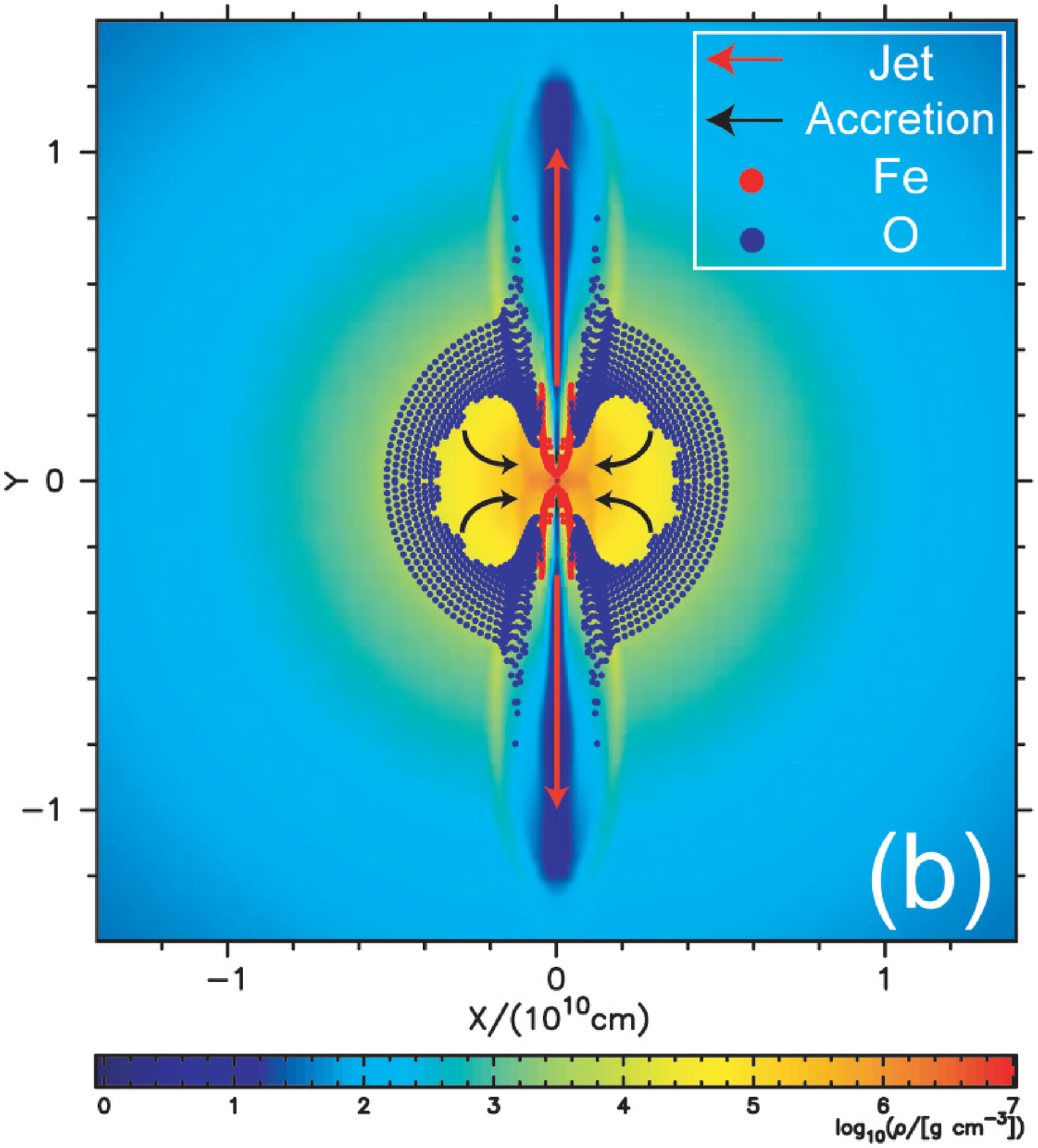} 
 \caption{(a) Schematic picture of the jet-induced explosion. (b) 
 Density structure of the 40 $\Msun$ Pop III star explosion model of
 $\Edep=15$ at 1 sec after the start of the jet injection. 
}
   \label{fig1}
\end{center}
\end{figure}

The jets are injected at a radius $\Rin$, corresponding to an
enclosed mass of $\Min$, and the jet propagation is followed
(Figs.~\ref{fig1}ab). Since the explosion mechanism is
unknown, the jets are
treated parametrically with the following five parameters: energy
deposition rate ($\Ed$), total deposited energy ($E_{\rm dep}$), initial
half angle of the jets ($\thj$), initial Lorentz factor
($\Gj$), and the ratio of thermal to total deposited
energies ($f_{\rm th}$).

In particular, we investigate the dependence of nucleosynthesis outcome
on $\Ed$ for a range of $\Edep\equiv\Ed/10^{51}{\rm ergs\,s^{-1}}=0.3-1500$.
The diversity of $\Ed$ is consistent with the wide range of the observed
isotropic equivalent $\gamma$-ray energies and timescales of 
GRBs (\eg \cite[Amati \etal\ 2007]{ama06}). 
Variations of activities of the central engines, possibly corresponding to
different rotational velocities or magnetic fields, may well produce 
the variation of $\Ed$. 
We expediently fix the other parameters as 
$\Et=1.5\times10^{52}$ergs, 
$\thj=15^\circ$, 
$\Gj=100$, 
$\fth=10^{-3}$, and $\Min=1.4\Msun$ ($\Rin\sim900$ km) in the models. 

The hydrodynamical calculations are followed until the homologously
expanding structure is reached ($v\propto r$) and then the ejected mass
elements are identified from their radial velocities. 
The nucleosynthesis calculations are performed as post-processing with
thermodynamic histories traced with marker particles that 
represent individual Lagrangian elements. 
In computing the jet composition, we assume that the jet initially has the
composition of the accreted stellar materials. 

\section{Jet-induced Supernovae}
\label{sec:result}

\begin{figure}
\begin{center}
 \includegraphics[width=2.in]{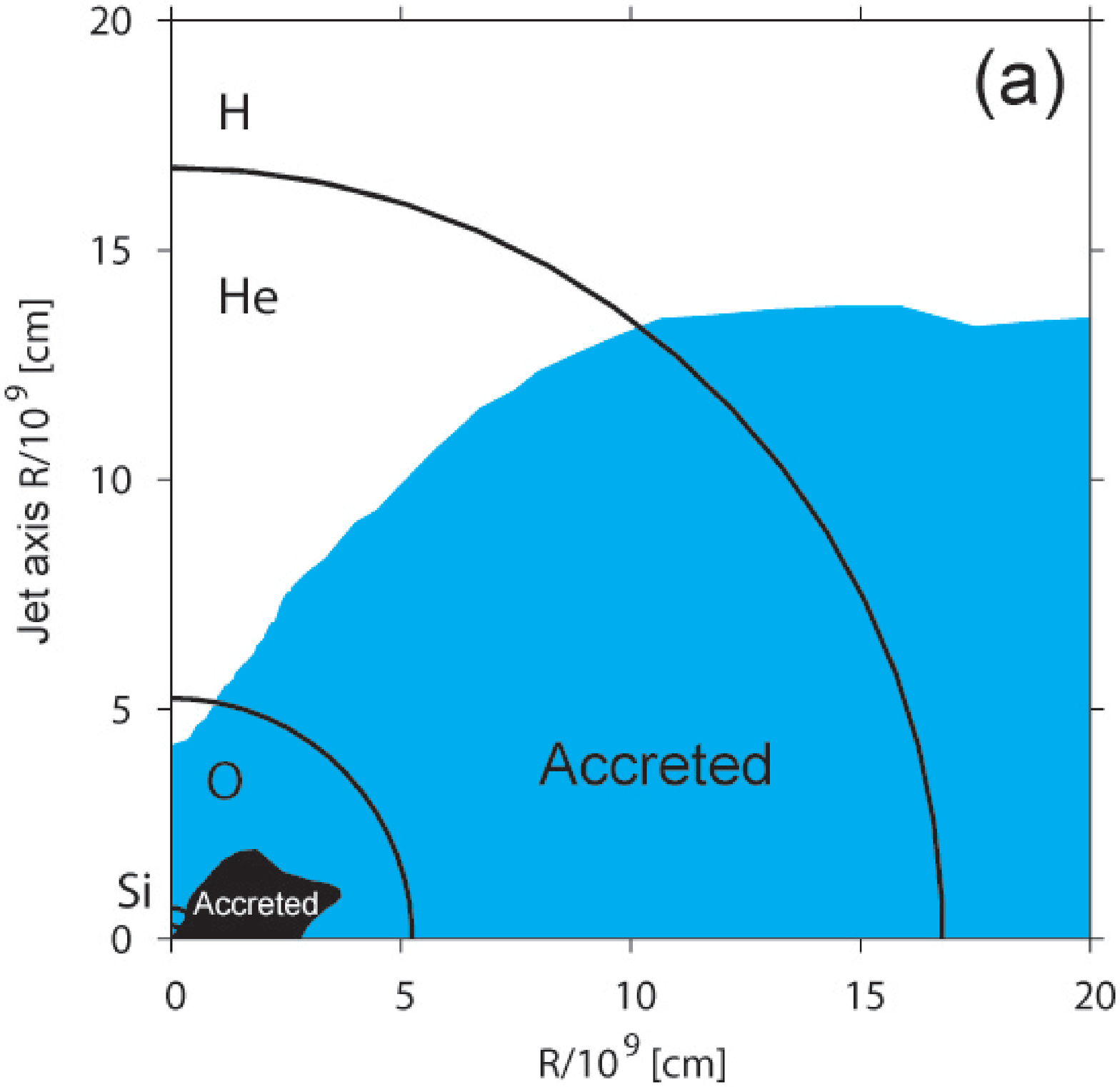} 
 \includegraphics[width=2.1in]{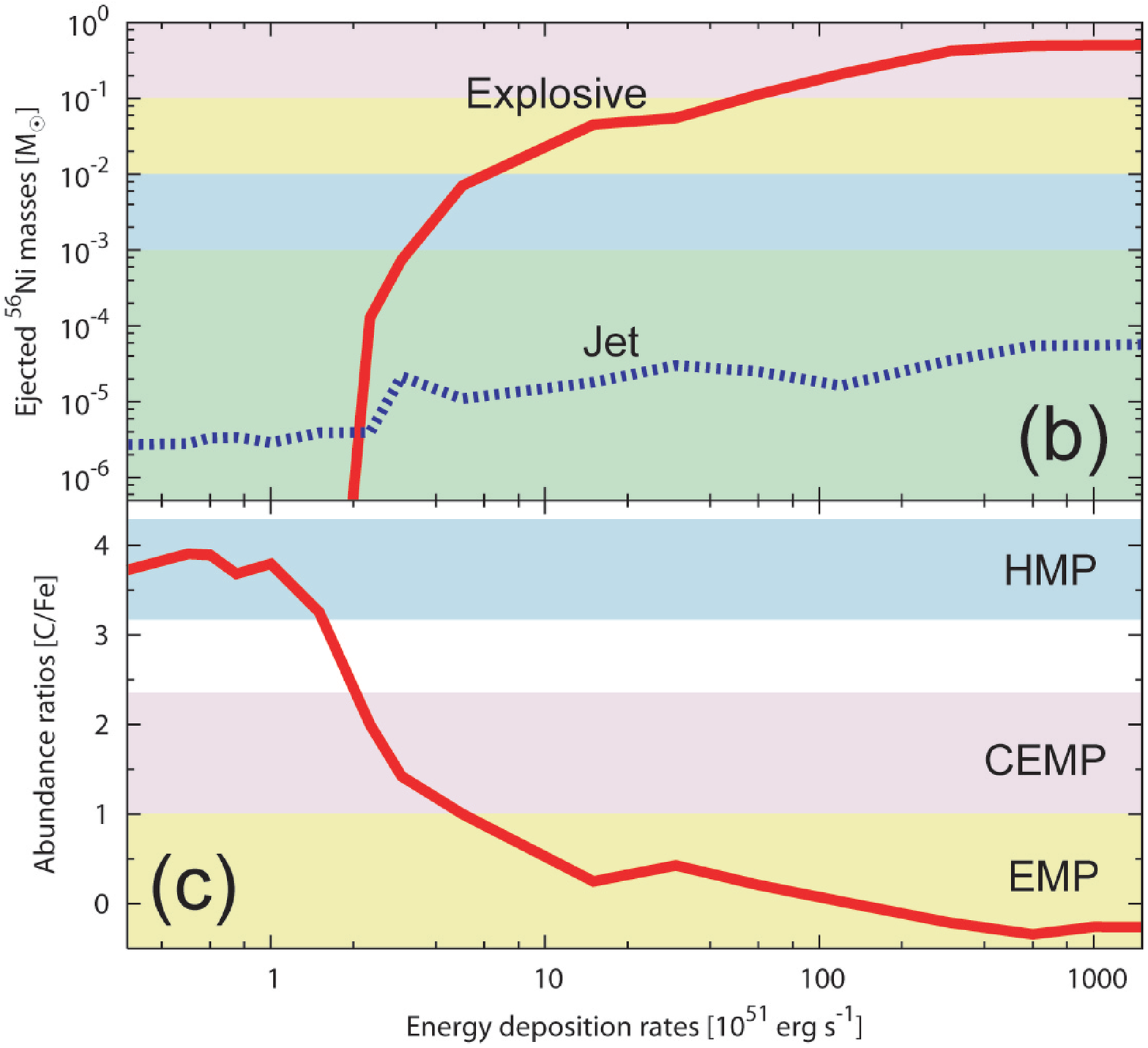} 
 \caption{(a) Initial locations of the mass elements which are finally
 accreted for models with $\Edep=120$ ({\it black}) and with $\Edep=1.5$
 ({\it cyan}). 
(b) Ejected Fe mass ({\it solid line}: explosive nucleosynthesis
 products, {\it dotted line}: the jet contribution)
 as a function of the energy deposition rate. 
(c) Dependence of abundance ratio [C/Fe] on the energy
 deposition rate. 
}
   \label{fig2}
\end{center}
\end{figure}

\subsection{Fallback}
\label{sec:fallback}

Figure~\ref{fig2}a show
``accreted'' regions for models with $\Edep=120$ and $1.5$ \ergs, where the accreted mass
elements initially located in the progenitor. 
The inner matter is ejected along the jet-axis but not along the
equatorial plane. On the other hand, the outer matter is ejected even
along the equatorial plane, since the lateral expansion of the shock
terminates the infall as the shock reaches the equatorial plane.

The remnant mass ($\Mbh$) is larger for lower $\Ed$. This stems from the balance
between the ram pressures of the injecting jet ($P_{\rm jet}$) and the
infalling matter ($P_{\rm fall}$). 
In order to inject the jet, $P_{\rm jet}$
should overcome $P_{\rm fall}$. 
$P_{\rm jet}$ is determined by $\Rin$, $\Ed$, $\thj$, 
$\Gj$, and $\fth$, thus being constant in time in
the present models. On the other hand, $P_{\rm fall}$ decreases with time,
since the density of the outer materials decreases following
the gravitational collapse (\eg \cite[Fryer \& M\'esz\'aros 2003]{fry03}). For lower $\Ed$,
$P_{\rm jet}$ is lower, so that the jet injection 
($P_{\rm jet}>P_{\rm fall}$) is realized at a later time when the
central remnant becomes more massive due to more infall. 
As a result, the accreted region and $\Mbh$ are larger for lower $\Ed$. 

A model with lower $\Ed$ has larger $\Mbh$, higher [C/Fe], and smaller
amount of Fe [$\Mfe$] because of the larger amount of fallback (Figs.~\ref{fig2}bc, 
\cite[Tominaga \etal\ 2007a]{tom07a}). 
The larger amount of fallback decreases the mass of the inner core
relative to the mass of the outer layer. The fallback of the O layer
also reduces $\Mfe$ because Fe is mainly synthesized explosively 
in the Si and O layers.
The variation of $\Ed$ in the jet-induced explosions predicts that
the variation of [C/Fe] corresponds to
that of $\Mfe$.

\subsection{Comparison with the spherical supernova model}
\label{sec:MFjet}

\begin{figure}
\begin{center}
 \includegraphics[width=1.8in]{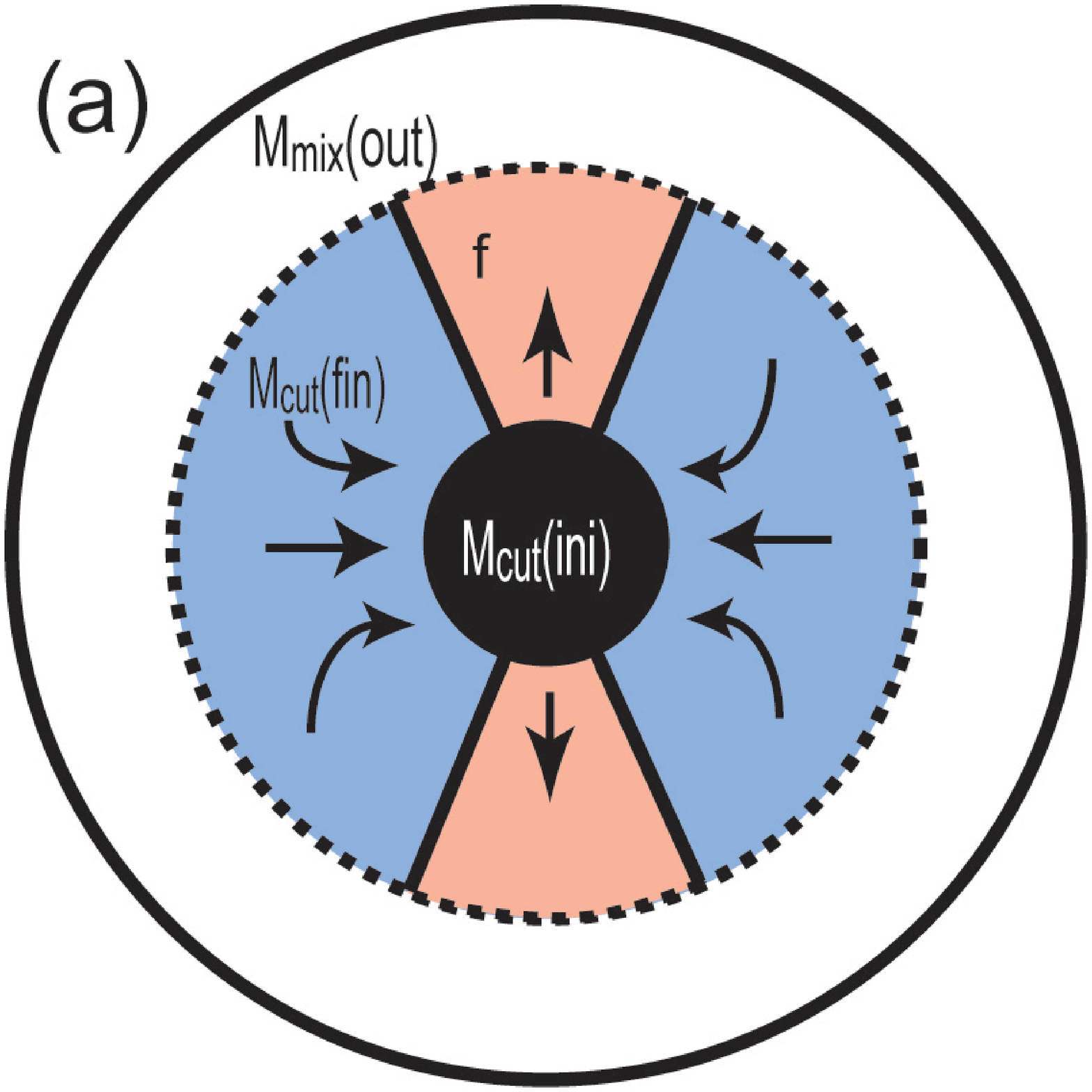} 
 \includegraphics[width=2.8in]{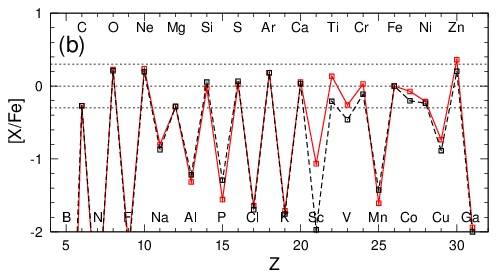} 
 \caption{(a) Two-dimensional illustration of the
mixing-fallback model. (b) Comparison of the abundance patterns of
 the jet-induced SN model with $\Edep=120$ and $\Min=2.3\Msun$ 
 ({\it solid line}) and the mixing-fallback
 model ({\it dashed line}).}
   \label{fig3}
\end{center}
\end{figure}

The calculations of the jet-induced explosions show that the ejection
of the inner matter is compatible with the fallback of the outer matter
(Fig.~\ref{fig2}a). This is consistent with the
two-dimensional illustration of the mixing-fallback model
(Fig.~\ref{fig3}a) proposed by \cite[Umeda \& Nomoto (2002)]{ume02}. 
The mixing-fallback model has three parameters; initial mass cut
[$M_{\rm cut}{\rm (ini)}$], outer boundary of the mixing region 
[$M_{\rm mix}{\rm (out)}$], and a fraction of matter ejected from the
mixing region ($f$). The remnant mass is written as 
\begin{equation}
 \Mbh=M_{\rm cut}{\rm (ini)}+(1-f)[M_{\rm mix}{\rm (out)} - M_{\rm cut}{\rm (ini)}].
\end{equation}
The three parameters relate to the hydrodynamical properties of
the jet-induced explosion.

A model with $\Edep=120$ and $\Min=2.3\Msun$ ($\Rin\sim3\times10^3$km) 
is compared with the spherical SN explosion adopting the mixing-fallback
model. The inner boundary and the central remnant mass of the jet-induced
explosion model are $\Min=2.3\Msun$ and $\Mbh=8.1\Msun$. 
Its abundance pattern is well reproduced by the 
spherical SN model with the same main-sequence mass $\Mms=40\Msun$ and
the explosion energy $E=3\times10^{52}$ ergs (Fig.~\ref{fig3}b). The parameters of the
mixing-fallback model are $M_{\rm cut}{\rm (ini)}=2.3\Msun$, 
$M_{\rm mix}{\rm (out)}=10.8\Msun$ and $f=0.19$.
The resultant $\Mbh$ ($=9.2\Msun$) is slightly larger than
$\Mbh$ of the jet-induce SN model.

There, however, are some elements showing differences. The differences
stem from the high-entropy explosion due to the
concentration of the energy injection (\eg \cite[Maeda \& Nomoto 2003]{mae03}). 
The enhancements of [Sc/Fe] and [Ti/Fe] improve agreements with the
observations. Such thermodynamical feature of the jet-induced explosion
model cannot be reproduced by the mixing-fallback model, while a
``low-density'' modification might mimic the high-entropy environment
(\eg \cite[Umeda \& Nomoto 2005; Tominaga \etal\ 2007b]{ume05,tom07b}).

\section{Trends with Metallicity}
\label{sec:trend}

The abundance patterns of the EMP stars show certain trends of abundance
ratios [X/Fe] with respect to [Fe/H] (\cite[Cayrel \etal\ 2004]{cay04}).
To clarify the origin of the trends, SN models with $\Mms=13-50\Msun$ are calculated. 
The explosion energies are set to be consistent with
the observations of present SNe (\eg \cite[Tanaka \etal\ 2008]{tan08}).

\begin{figure}
\begin{center}
 \includegraphics[width=4.2in]{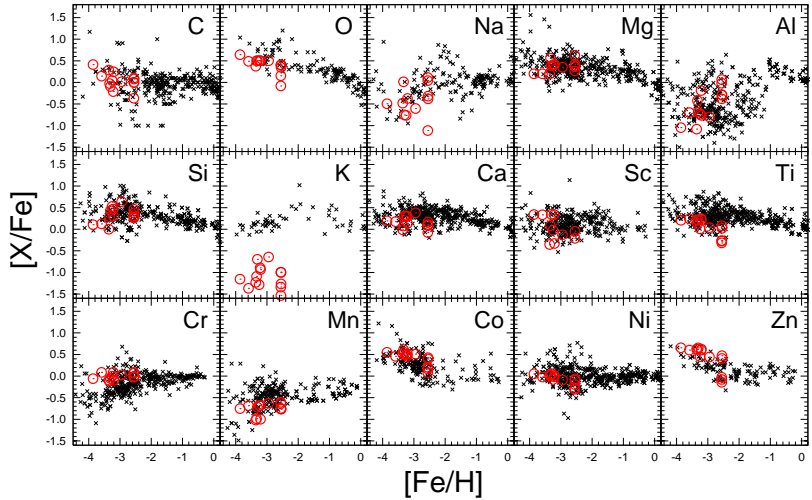} 
 \caption{Comparison between the [X/Fe] trends of observed stars
 (\eg \cite[Cayrel \etal\ 2004; Honda \etal\ 2004]{cay04,hon04}: {\it cross}) and SN models with the
 mixing-fallback model and applied the $Y_{\rm e}$ and ``low-density''
 modifications (\cite[Tominaga \etal\ 2007b]{tom07b}: {\it circles}). }
   \label{fig4}
\end{center}
\end{figure}

The abundance ratios against
[Fe/H] are compared with yields of individual SN models and the
Salpeter's IMF-integrated yield (Fig.~\ref{fig4}). [Fe/H] of a
next-generation star is determined by $\Mfe$ and a swept-up H mass. 
Since the swept-up H mass is almost proportional to $E$
of the SN, 
[Fe/H] of the next-generation stars are determined by a equation 
${\rm [Fe/H]} = \log_{10}\left[{{M({\rm Fe})\over{\Msun}}/{\left({E\over{10^{51}{\rm ergs}}}\right)^{6/7}}}\right]-C$  
(\cite[Thornton \etal\ 1998]{tho98}), 
where $C$ is assumed to be a constant value of 1.4.
[Fe/H] of the IMF-integrated abundance
ratios are assumed to be the same as SN models with $E\sim10^{51}$ergs ([Fe/H]~$\sim -2.6$).
 
HNe explode with $E\gsim10^{52}$ergs and eject large amount of
\Nifs\ ($\gsim0.1\Msun$), while normal SNe explode with $E\sim10^{51}$ergs and
eject $\sim0.07\Msun$ of \Nifs. Therefore, according to the above equation,
[Fe/H] of a next-generation star originated from a HN is lower than that of a
next-generation star originated from a normal SN. The higher-energy explosion
raises explosive nucleosynthesis under higher entropy and thus leads
higher [Zn/Fe]. This accompaniment explains the observed trends of
[Zn/Fe] against [Fe/H].
The trends of other elements are also reproduced by the variations of
$\Mms$ and $E$ (Fig.~\ref{fig4}).

\section{Conclusion}

We focus on two interesting properties observed in the abundance patterns of the
metal-poor stars: (1) the higher
[C/Fe] for lower [Fe/H] and (2) the trends of [X/Fe]
against [Fe/H]. The variations of the metal-poor stars are explained by
the variations of SNe that contribute the metal enrichment of the early
universe. Especially, (1) the variation of the energy deposition rates
explains the tendency of [C/Fe] against [Fe/H] and (2)
the variations of $\Mms$ and $E$
explain the trends of [X/Fe] against [Fe/H]. We propose that the
abundance patterns of the metal-poor stars will provide additional
constraints on the explosion mechanism of GRBs and SNe other than the
direct observations of present GRBs and SNe.

\vspace{.5cm}

\noindent This work has been supported in part by WPI Initiative, MEXT, Japan.


\begin{thebibliography}{}

\bibitem[Amati \etal (2007)]{ama06} Amati, L., Della Valle, M.,
			   Frontera, F., \etal\ 2007,
			   \textit{A\&A}, 463, 913

\bibitem[Beers \& Christlieb(2005)]{bee05} Beers, T.C., \& Christlieb,
			   N. 2005, \textit{ARAA}, 43, 531

\bibitem[Cayrel \etal(2004)]{cay04}Cayrel, R., Depagne, E., Spite, M., \etal\ 2004,
			   \textit{A\&A}, 416, 1117 

\bibitem[Depagne \etal (2002)]{dep02} Depagne, E., Hill, V., Spite, M., 
			    \etal\ 2002, \textit{A\&A},
			   390, 187

\bibitem[Frail \etal (2001)]{fra01} Frail, D.A., Kulkarni, S. R., Sari,
			    R., \etal\ 2001, \textit{ApJ} (Letters), 562, L55

\bibitem[Fryer \& M\'esz\'aros(2003)]{fry03} Fryer, C., \& M\'esz\'aros,
			   P. 2003, \textit{ApJ} (Letters), 588, L25

\bibitem[Galama \etal(1998)]{gal98} Galama, T. J., Vreeswijk, P. M., van
			    Paradijs, J., \etal\ 1998,
\textit{Nature}, 395, 670

\bibitem[Honda \etal(2004)]{hon04} Honda, S., Aoki, W., Kajino, T.,
			    \etal\ 2004, \textit{ApJ}, 607, 474

\bibitem[Iwamoto \etal(2005)]{iwa05} Iwamoto, N., Umeda, H., Tominaga,
				N., Nomoto, K., \& Maeda, K. 2005,
				\textit{Science}, 309, 451 

\bibitem[Maeda \etal(2002)]{mae02} Maeda, K., Nakamura, T., Nomoto, K.,
			     \etal\ 2002, \textit{ApJ}, 565, 405

\bibitem[Maeda \& Nomoto (2003)]{mae03} Maeda, K., \& Nomoto, K. 2003,
			   \textit{ApJ}, 598, 1163

\bibitem[Nagataki \etal(2006)]{nag06} Nagataki, S., Mizuta, A., \& Sato,
			   K. 2006, \textit{ApJ}, 647, 1255

\bibitem[Nomoto \etal (2006)]{nom06} Nomoto, K., Tominaga, N., Umeda, H.,
				\etal\ 2006,
			   \textit{Nucl. Phys. A}, 777, 424 (astro-ph/0605725)

\bibitem[Tanaka \etal(2008)]{tan08} Tanaka, M., Tominaga, N., Nomoto,
			   K., \etal\ 2008, \textit{ApJ}, submitted
			   (arXiv:0807.1674)

\bibitem[Thornton \etal(1998)]{tho98} Thornton, K., Gaudlitz, M., Janka,
			   H.-Th., \& Steinmetz, M. 1998, \textit{ApJ}, 500, 95

\bibitem[Tominaga \etal(2007a)]{tom07a} Tominaga, N., Maeda, K., Umeda,
			   H., \etal\ 2007a, \textit{ApJ} (Letters), 657,
			   L77

\bibitem[Tominaga \etal(2007b)]{tom07b} Tominaga, N., Umeda,
			   H., \& Nomoto, K. 2007b, \textit{ApJ}, 660, 516

\bibitem[Tominaga(2009)]{tom07c} Tominaga, N. 2009, \textit{ApJ},
			   in press (arXiv:0711.4815)

\bibitem[Umeda \& Nomoto(2002)]{ume02} Umeda, H., \& Nomoto, K. 2002,
			    \textit{ApJ}, 565, 385 

\bibitem[Umeda \& Nomoto(2005)]{ume05} Umeda, H., \& Nomoto, K. 2005,
\textit{ApJ}, 619, 427

\bibitem[Woosley \& Weaver(1995)]{woo95} Woosley, S. E., \& Weaver,
			     T. A. 1995, \textit{ApJS}, 101, 181

\end{thebibliography}
\end{document}